\documentclass[%
reprint,
longbibliography,
amsmath,amssymb,
aps,
]{revtex4-1}

\usepackage{graphicx}
\usepackage{dcolumn}
\usepackage{bm}
\usepackage{xcolor}

\renewcommand{\bf}{\textbf}

\newcommand{\E}{\varepsilon}

\def\e{\begin{equation}}
\def\f{\end{equation}}
\def\_#1{{\bf #1}}
\def\=#1{\overline{\overline #1}}

\def\E{\varepsilon}

\def\.{\cdot}

\begin{document}
	
	\preprint{APS/123-QED}

\title{Brewster effect when approaching exceptional points of degeneracy:\\ Epsilon-near-zero behavior }
		\author{Vladislav Popov}
	\email{uladzislau.papou@centralesupelec.fr}
	\affiliation{%
		SONDRA, CentraleSup\'elec, Universit\'e Paris-Saclay,
		F-91190, Gif-sur-Yvette, France
	}%
    \author{Sergei Tretyakov}
	\email{sergei.tretyakov@aalto.fi}
	\affiliation{%
		Aalto University, P.O. 15500, FI-00076 Aalto,
     Finland
     	}%
	\author{Andrey Novitsky}%
	\email{anov@fotonik.dtu.dk}
	\affiliation{%
		DTU Fotonik, Technical University of Denmark, {\O}rsteds Plads 343, DK-2800 Kongens Lyngby, Denmark
	}%
    \affiliation{%
		Department of Theoretical Physics and Astrophysics, Belarusian State
 		University, Nezavisimosti av. 4, 220030 Minsk, Belarus
	}%

\begin{abstract}
We reveal that the phenomenon of full transmission without phase accumulation commonly associated with epsilon-near-zero (ENZ) materials for a plane-wave  does  not require  vanishing of  permittivity. 
We theoretically connect the phenomenon with condition of the Brewster effect satisfied at the edges of stop bands (so called exceptional points of degeneracy) and show that the full transmission without phase accumulation can be observed in various one-dimensional periodic structures. 
Particularly, exploiting the manifold of exceptional points of degeneracy in one-dimensional all-dielectric periodic lattices, we demonstrate that these structures not only offer a lossless and extremely simple, CMOS compatible alternative for some applications of ENZ media, but exhibit new properties of all-angle full transmission with zero phase delay.
\end{abstract}

\maketitle

Metamaterials promise amazing possibilities in manipulation of electromagnetic fields, which are not available with natural materials (see, e.g.,   \cite{Jacob2016,Alu2016,Glybovski2016,C7TC03384B,Tong2018}). Extreme properties and unique effects leading to novel functionalities require artificial materials with extreme and singular values of material parameters. Recently, much attention has been attracted by exotic properties of materials with permittivity $\varepsilon$   near zero. Both natural substances and metamaterials possessing $\varepsilon \approx 0$ are called epsilon-near-zero (ENZ) media. From the physical point of view, this special value of permittivity corresponds to the topological transition between metals and dielectrics, as  can be illustrated by tuning parameters of a hyperbolic metamaterial, whose wave dispersion changes from hyperbolic to elliptic type \cite{Poddubny2013,Drachev2013} at the ENZ point. From the applications point of view, properties of  ENZ media can be exploited for  energy tunneling through subwavelength channels~\cite{PhysRevLett.100.023903,PhysRevLett.100.033903}, improvement of antenna directivity~\cite{PhysRevB.75.155410}, phase matching due to zero phase advance in ENZ media~\cite{PhysRevB.89.235401}, enhancement of nonlinear effects~\cite{PhysRevA.84.063826,PhysRevB.85.045129}, 
etc. \cite{Feng2012,Liberal2017,LiberalSci2017}. 

%
\begin{figure}[tb]
\includegraphics[width=0.8\linewidth]{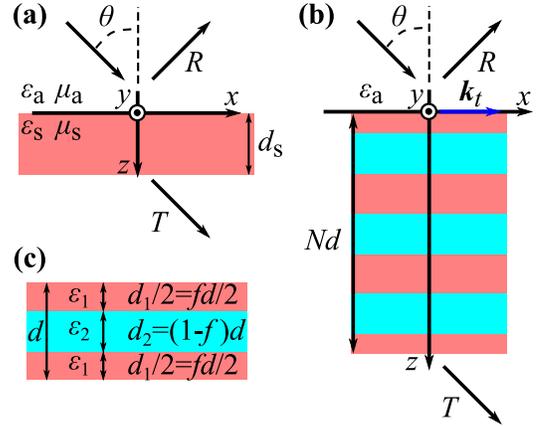}
\caption{\label{fig:0} (a) Reflection and transmission through an isotropic slab and (b) 1D photonic crystal composed of alternating layers of isotropic dielectrics. (c) Three-layer (having an inversion center) unit cells of a 1D photonic crystal.}
\end{figure}

Conditions for $\varepsilon \approx 0$ have been the subject of scrutiny for more than a decade \cite{Garcia2002,Liberal2017}.
Unfortunately, natural ENZ materials are rather lossy, and this may suppress or even ruin their useful properties \cite{Javani2016}. 
To circumvent this difficulty  one may exploit the accidental degeneracy near the $\Gamma$-point in all-dielectric photonic crystals, where the zero-refractive-index properties are available \cite{huang2011dirac,moitra2013realization}. On the other hand, superlattices constructed of positive and negative index photonic crystals can also be  used~\cite{Panoiu:06,kocaman2011zero}. 
However, it is important to keep in mind that all artificial ENZ materials are complex 2D or 3D lattices that can be characterized by effective permittivity only under the homogenization condition (period is much smaller than the operating wavelength).

In this letter, we show that a physical mechanism behind properties of ENZ media does not require vanishing of permittivity. We theoretically connect the effect of complete wave tunneling without phase accumulation to the condition of the Brewster effect  satisfied when approaching exceptional points of degeneracy.  This theory opens up a possibility to exploit the manifold of exceptional points of degeneracy in all-dielectric periodic lattices in order to emulate properties of ENZ media. Particularly, we demonstrate that one-dimensional (1D) periodic structures can be designed in order to exhibit ENZ behavior. We validate the theory by demonstrating the complete wave tunneling without phase advance through simple layered  structures composed of conventional dielectrics. We also show that there is no phase accumulation throughout the whole thickness of a structure.


To explain the analogy between existence of exceptional points of degeneracy, the Brewster effect,  and the properties of ENZ media, let us first consider the well-known analytical solution for the plane wave transmission through an isotropic slab of thickness $d_s$, permittivity $\E_s$, and permeability $\mu_s$, as illustrated in Fig.~\ref{fig:0}(a). The transmission coefficient for an obliquely incident plane-wave of arbitrary polarization can be written as~\cite{PhysRevB.75.155410}
\begin{equation}\label{eq:T_slab}
T=\left[\cos(k_{zs}d_s)-i\frac{Z_s^2+Z_a^2}{2Z_sZ_a}\sin(k_{zs}d_s)\right]^{-1}.
\end{equation} 
Here, $k_{zs} = \sqrt{\varepsilon_s \mu_s k_0^2 - k_t^2}$ is the normal component of the wave vector inside the slab, $k_0=\omega/c$ is the vacuum wavenumber ($\omega$ is the frequency, $c$ is the speed of light in  vacuum), $k_t$ represents the tangential component of the wave vector,  $Z_s$ and $Z_a$ are the wave impedances  inside and outside  the slab, respectively. 
The impedances are defined as the ratios of the tangential to the slab plane-wave field components.
Assuming absence of losses, the reflectivity $1-|T|^2$ vanishes provided $Z_s=Z_a$. 
Except for the trivial case of the same materials of the slab and the ambient medium, the equal wave impedances are realized under conditions of Brewster's law. 
In this case the transmission coefficient $T=e^{ik_{zs}d_s}$ describes the phase shift of the fully transmitted wave.  When the full transmission is complemented  by  zero phase shift $k_{zs}d_s = 0$, a unique  phenomenon of complete wave tunneling without phase accumulation is realized. Originally, it was associated with the ENZ/mu-near-zero(MNZ)/epsilon-mu-near-zero(EMNZ) materials. 

Noteworthy, a Fabry-Perot  resonance occurring in any dielectric slab under the conditions $k_{zs}d_s=2\pi m$ ($m=1,2,\ldots$) is also characterized by the full transmission and zero phase shift.
However, the Brewster effect is fundamentally different from the Fabry-Perot resonance. It exists at any thickness of the slab $d_s$. Moreover, a plane wave does not accumulate phase when it propagates across the slab under conditions $Z_s=Z_a$ and $k_{zs}=0$. On the output the phase accumulation is $0$, but not a multiple of $2\pi$.

\begin{figure}[tb]
\includegraphics[width=0.99\linewidth]{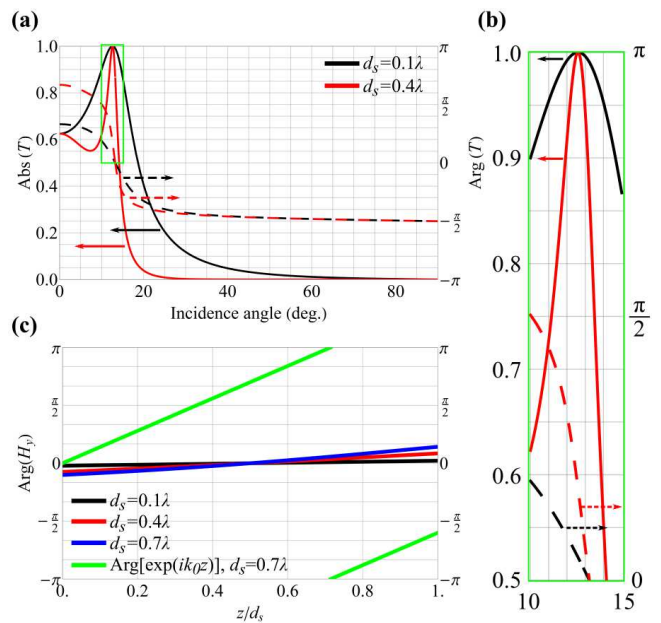}
\caption{\label{fig:1} (a) Transmission coefficient $T$ through a dielectric slab for TM polarization versus incidence angle. (b) Zoom of the boxed area in (a).  (c) The phase distribution inside the slab under  the critical angle incidence. The green line describes the phase evolution  of a plane-wave propagating through vacuum within the distance  $d_s$. In all figures $\lambda$ represents the vacuum wavelength. Parameters: $\E_a=20$, $\mu_a=1$, $\E_s=1$ and $\E_s=1$. The Brewster angle and the critical angle of TIR equal  $12.6^\circ$ and $12.9^\circ$, respectively. }
\end{figure}
\begin{figure*}[tb]
	\includegraphics[width=0.99\linewidth]{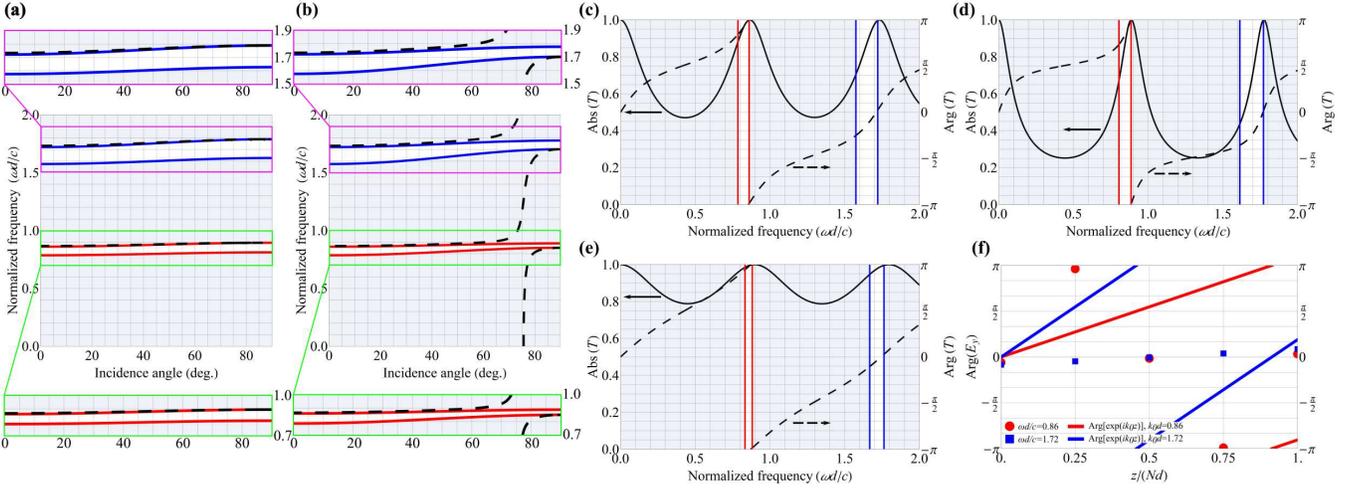}
	\caption{\label{fig:2} Band-gap structure for (a) TE and (b) TM polarizations of a 1D photonic crystal depicted in Fig~\ref{fig:0} (b). The black dashed curves correspond to Brewster's law. (c)--(e) Absolute value and phase of the transmission coefficient through a single unit cell of the photonic crystal versus the normalized frequency: (c) normal incidence, (d) TE polarization, $\theta=60^\circ$, (e) TM polarization, $\theta=60^\circ$. In all figures shaded regions represent propagation bands and  red and blue curves and lines represent the edges of stop bands where $q=\pi/d$ and $q=0$, respectively. (f)  Spatial evolution of the electric field phase over periods of the $4$-cell 1D photonic crystal (circles and cubes) compared to the phase accumulation of a plane wave propagating through a vacuum region of equivalent thickness (solid lines). Used parameters are $\E_1=2$, $\E_2=16$ and $f=0.1$.}
\end{figure*}
%

%
Condition $k_{zs} = 0$ defines the critical angle $\theta_c$ of the total internal reflection (TIR). Here we utilize only  the critical angle condition $k_{zs} = 0$ while the phenomenon of TIR itself is not observed for the considered slabs, being available only for very thick layers.
Since we need the two conditions $Z_s = Z_a$ and $k_{zs} = 0$ to be satisfied simultaneously for full transmission without phase accumulation, the Brewster angle and the critical angle of total internal reflection should be close.
%
%
Indeed, as illustrated by Fig.~\ref{fig:1}, the TM-polarized wave is fully transmitted with no phase accumulation for conventional dielectric materials ($\E_a=20$ and $\E_s=1$). Even though the Brewster angle ($\theta_B=12.6^\textup o$) and the critical angle of TIR ($\theta_c=12.9^\textup o$) do not coincide exactly, the system is highly transparent at $\theta=\theta_c$, see Fig.~\ref{fig:1} (b). Similarly to an ENZ case the phase accumulation across the slab is insignificant unlike the unbounded plane-wave propagation in vacuum as shown in Fig.~\ref{fig:1} (c).
Interestingly, the investigated phenomenon does not suffer from  frequency dispersion, because the critical angle of TIR and Brewster's angle depend only on the practically dispersionless material parameters of the dielectrics, but not on the frequency. It is demonstrated in  Fig.~\ref{fig:1} (b) with the help of two curves corresponding to different thicknesses $d_s$ of the slab.
Thus, we have obtained surprising results demonstrating that the permittivity should not necessarily vanish to achieve  high transmission without phase accumulation. Instead, under specific conditions, conventional dielectrics can be exploited.

Generally speaking, the critical angle of TIR corresponds to an exceptional point of degeneracy (EPD) since the condition $k_{zs}=-k_{zs}=0$ means that the propagation constants of the oppositely directed waves coincide (degenerate), while the eigenvectors and, hence, eigenwaves' impedances are indistinguishable. 
Thus, the phenomenon of full transmission without phase accumulation  can be found for many different structures. 
Indeed, let us consider a simple (for analytical derivations) example  of a 1D periodic structure which can be described by an ABCD matrix~\cite{pozar2009microwave} (or, in other words, transfer matrix) and has a  symmetric unit cell (in this case $A=D$~\cite{pozar2009microwave}) of geometric size $d$. 
The asymmetric case ($A\neq D$) is discussed in the Supplementary Material~\cite{spl}. 
Bloch waves characterized by a wavenumber $q$ and wave impedance $Z_B$ propagate through the periodic structure. Keeping in mind the unimodularity of the ABCD matrix ($A^2-BC=1$) for the non-dissipative structure, one can write the Bloch wavenumber $q=\pm d^{-1}\cos^{-1}(A)$ and wave impedance $Z_B=\pm\sqrt{C/B}$ solving the eigenvalue and eigenvector problem for the ABCD matrix~\cite{pozar2009microwave}.
Here signs plus and minus  correspond to forward and backward Bloch waves. They cannot be distinguished at the edges between stop and pass bands (i.e. when $q=0$ and $\pi/d$), where the wave impedance diverges or vanishes by means of $B=0$ or $C=0$.
Thus, the edges of stop bands can be always treated as EPDs.
Transmission coefficient through a $N$-cell 1D periodic structure is given by the same Eq.~\eqref{eq:T_slab}, but the wave impedance of the Bloch wave $Z_B$, Bloch wavenumber $q$, and $Nd$ should substitute respectively $Z_s$, $k_{sz}$, and $d_s$. When the Brewster condition $Z_B=Z_a$ is satisfied close to a band gap edge, the transmission coefficient $T=\exp(iqNd)$ is able to approach unity arbitrarily close.
As in the case of a homogeneous ENZ slab, the wave does not accumulate  phase when propagating. Thus, a discrete analogue of the wave propagation phenomenon in homogeneous ENZ media can be achieved in periodic structures. To that end one needs to appropriately design a unit cell to get a required ABCD matrix, e.g., see Ref.~\cite{pozar2009microwave}, where a design of microwave networks is discussed.

As a concrete example we consider a 1D photonic crystal
represented by a periodic multilayer composed of three-layer unit cells with an inversion center~\cite{Popov2018_OEMA_SW}, as shown in Figs.~\ref{fig:0}(b) and (c). EPDs are well studied in such structures, see, e.g., \cite{PhysRevE.53.4107,PhysRevB.67.165210,PhysRevE.72.036619}. 
Complete wave tunneling without phase accumulation in 1D photonic crystals is available for any incidence angle $\theta$ and any polarization, if the boundaries of the stop bands are almost flat.
The required band structure can be achieved for alternating low- and high-permittivity dielectric layers with a small filling fraction $f = d_1/d$ of the low-permittivity dielectric. Then the stop bands become narrower, while their edges are almost flat for both TE  [Fig.~\ref{fig:2} (a)] and TM polarizations [Fig.~\ref{fig:2} (b)].
%
%
The full transmission $|T|=1$ independent of the total thickness $Nd$ (Brewster's resonance) appears at $Z_B=Z_a$. 
The values of the incidence angles and frequencies at which Brewster's law in periodic multilayers is satisfied are depicted as dashed lines in Figs.~\ref{fig:2} (a) and (b). It should be noticed that Bloch impedance $Z_B$ is frequency dependent what  gives rise  to frequency dispersion.
For  small filling fractions $f$, the curves $Z_B=Z_a$ shift to the top edges of the stop bands, thus, providing both full transmission and zero phase accumulation.
For a discussion on EPDs in the considered 1D photonic crystals 
see Supplementary Material~\cite{spl}.

Transmission coefficient through a  single unit cell ($N=1$) as a function of the normalized frequency $k_0d=\omega d/c$ is demonstrated in Figs.~\ref{fig:2} (c)--(e).   
All maxima in Figs.~\ref{fig:2} (c) -- (e) correspond to Brewster's resonances which are independent of the total thickness $Nd$. 
There are no Fabry-Perot resonances at the  band gap edges $q=0$ and $q=\pi/d$ in Figs.~\ref{fig:2} (c)--(e), since $Z_B$ goes either to zero or infinity, while the term $(Z_B^2+Z_a^2)/(2Z_BZ_a)\sin(qNd)$ takes a non-zero value and, therefore, $|T|<1$ according to Eq.~\eqref{eq:T_slab}. 
On the contrary, when the Brewster condition $Z_B = Z_a$ is satisfied next to stop band edges, transmission coefficient can approach $unity$ arbitrarily close when sufficiently small $f$ is chosen (of course, $f$ cannot be exactly zero as there would be no band gap). 
%
Except for the case of  normal incidence $\theta = 0$, the multilayer structure is polarization-sensitive. 
When the incidence angle of the TE(TM) wave increases, the stop bands get wider (narrower) [see Figs~\ref{fig:2} (a) and (b)], while the widths of the transmission resonances shown in Figs~\ref{fig:2} (c)-(f) decrease (increase).  
Transmission coefficients for the waves at the top edges of stop bands $q=\pi/d$ and $q=0$ possess the phases respectively $\pi$ or $0$ as it is shown by dashed curves in Figs.~\ref{fig:2} (c)--(e). 
\begin{figure}[tb]
	\includegraphics[width=0.99\linewidth]{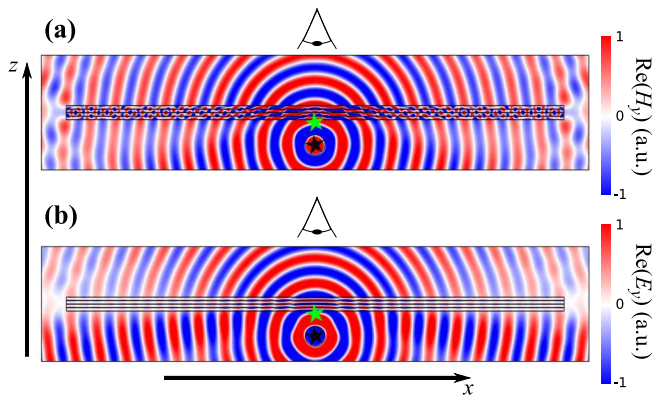}
	\caption{\label{fig:4} (a) Distribution of the magnetic field when a magnetic line source illuminates a 1D photonic crystal. (b) Distribution of the electric field when an electric line source illuminates the photonic crystal. Normalized  frequency is $k_0d=0.86$, the distance to the photonic crystal is $\lambda$, the 1D photonic crystal has $N=4$ three-layer unit cells and the length of $20\lambda$. The eye represents an observer who sees an illusion (top star) of the source (bottom star).}
\end{figure}
%

Figure ~\ref{fig:2} (f) demonstrates the phase of the electric field at the points multiple to the period of the $4$-cell multilayer. At the top edge of the first stop band $q=\pi/d$ the phases are close for even and odd periods, and  there is no phase accumulation at discrete points $z_m = 2 m d$ and $z_m = (2m+1) d$ ($m$ is an integer number). At the top edge of the second stop band $q=0$ the phases are nearly equal after every period of the multilayer, i.e. there is no phase accumulation at discrete points $z_m = m d$. To compare, plane waves propagating in unbounded vacuum of the same thickness experiences significant phase accumulation as shown by the solid lines in Fig.~\ref{fig:2} (f). We should stress  that the phase within each unit cell chandes significantly, albeit the phase accumulation between the periodically arranged points is almost absent. See Supplementary Material~\cite{spl} for more details on the spatial evolution of the phase. 

Realistic losses and random fluctuations of geometrical parameters of the dielectric structures do not affect much the overall performance (see Supplementary Material~\cite{spl} for  details) in bright contrast to the ENZ media~\cite{Javani2016}. 
Additionally, it is worth to note that we have considered only propagation of monochromatic plane waves, however, an information carrying signal  would consists of a spectrum of such waves. From Fig.~\ref{fig:2} one can get information about the frequency response and, particularly, see that the group delay $\partial\textup{Arg}(T)/\partial\omega$ (an important characteristic in signal processing~\cite{Oppenheim1996}) increases when approaching the frequencies of the top  edges of stop bands (in coherence with the well-known slow light phenomenon~\cite{Dowling1994}). This observation brings us to conclusion that only spectrally narrow signals can propagate without distortion through the considered photonic crystal when showing ENZ properties. See Supplementary Material for more details.


The polarization insensitive illusion effect represents an impressive demonstration of the discussed phenomenon. An ideal device making an illusion is invisible; therefore, its realization using a 1D photonic crystal, which completely tunnels the wave without a phase advance in a vast range of incident angles, is natural.
In Fig.~\ref{fig:4}, a source is placed below the photonic crystal and radiates at the frequencies specified by the top edge of the stop band. Then an observer sees the source at a shorter distance, the displacement towards the observer being equal to the thickness of the photonic crystal slab. Such an illusion is demonstrated in Fig.~\ref{fig:4} for magnetic and electric line sources, i.e. for TE and TM polarizations.
It is worth noticing that an illusion phenomenon was reported in Refs.~\cite{PhysRevLett.102.253902,doi:10.1063/1.3371716,6922484,Yao:17} on the base of transformation optics. 

To summarize, we have identified the physical mechanism behind the properties of ENZ media, namely, the Brewster effect occurring next to exceptional points of degeneracy of an electromagnetic structure. This has  allowed us to demonstrate that vanishing of permittivity is not required for achieving full transmission without phase accumulation and can be realized with many different 1D periodic structures. Particularly, we demonstrated the phenomenon using simple layered structures composed of conventional dielectrics. These structures not only offer a lossless and extremely simple, CMOS compatible alternative for some applications of ENZ media, but exhibit new properties of all-angle polarization insensitive full transmission with zero phase accumulation. Weak  sensitivity of the wave tunneling property to variations of electromagnetic and geometrical parameters of the structure makes this system attractive for realization of such extreme effects as source-displacement illusion. 
Importantly, since exceptional points of degeneracy are ubiquitous, we expect that one can also realize ENZ-like properties with more complex 2D and 3D non-homogenizable structures.

\newpage

\section*{Supplementary Material}

\section*{S1: ENZ behavior in 1D periodic structures with asymmetric unit cell}
\label{app:a}

Electromagnetic wave propagation through one-dimensional periodic structures can be described by means of the ABCD matrix approach~\cite{pozar2009microwave}. In the main text, we discuss conditions of full transmission without phase accumulation for waves penetrating periodic structures with a symmetric unit cell. Here we focus on the distinct case of an asymmetric unit cell resulting in $A\neq D$.  Then, a Bloch wavenumber $q$ is found to be~\cite{pozar2009microwave}
\begin{equation}\label{eq:q_general}
q=\pm\frac{1}{d}\cos^{-1}\left[\frac{A+D}{2}\right].
\end{equation}
Meanwhile, wave impedances of the eigenwaves are given by the equation
\begin{equation}\label{eq:WI_ASL}
Z_{B}^\pm=\frac{C}{\frac{A-D}{2}\mp i\sin(qd)}.
\end{equation}
The top and bottom sings correspond to the waves propagating along and against $z$-axis, respectively.
From the formula~\eqref{eq:WI_ASL} it follows that $Z_B^+\neq-Z^-_B$
in this general case. 
Furthermore, even in the lossless scenario the wave impedance $Z_{B}^\pm$ is complex-valued contrary to the case of symmetric unit cells when the impedance is either real (passband band) or imaginary (stop band).
Then, the formula~(1) for the transmission coefficient takes the following, more general, form
\begin{equation}\label{eq:T_gen}
T=\left[\cos(qNd)-i\frac{Z_a^2-Z^+_BZ^-_B}{Z_a(Z_B^+-Z_B^-)}\sin(qNd)\right]^{-1}.
\end{equation}
Let us consider the behavior of the transmission coefficient at a stop band edge.
It may seem that  the transmission coefficient equals $1$ when, e.g., $q=0$. However, when approaching  a boundary of a stop band,  $\sin(qd)\rightarrow 0$ and accordingly $Z_B^+\rightarrow Z_B^-$. That is, wave impedances at a stop band boundary are not independent while the expression for the difference $(Z_B^+-Z_B^-)$ can be reduced as follows:
\begin{eqnarray}\label{eq:difference}
&&Z_B^+-Z_B^-=
\frac{C}{\frac{A-D}{2}- i\sin(qd)}-\frac{C}{\frac{A-D}{2}+ i\sin(qd)}\nonumber\\
&&=\frac{2Ci\sin(qd)}{\left(\frac{A-D}{2}\right)^2+ \sin(qd)^2}.
\end{eqnarray}
As a result, the denominator of the fraction $(Z_a^2-Z^+_BZ^-_B)/[Z_a(Z_B^+-Z_B^-)]$ goes to zero while the numerator has a nonzero limit. Since the difference $(Z_B^+-Z_B^-)$ is proportional to $\sin(qd)$ [when $\sin(qd)$ is close to zero, of course], $\sin(qNd)/(Z_B^+-Z_B^-)$ also has a nonzero limit proportional to $N$. Eventually, the transmission coefficient does not go to $1$ at a stop band boundary. 
However, if $\sqrt{Z^+_BZ^-_B}$  goes to $Z_a$  when approaching a stop band boundary one would observe full transmission without phase accumulation.

\section*{S2: Additional details on the ENZ behavior of 1D periodic structures with a symmetric unit cell}
\label{app:b}

In this section, we provide additional details (on the example of a 1D photonic crystal) on the exceptional points of degeneracy in  periodic structures having a symmetric unit cell. After that we show a limiting behavior of the transmission coefficient at stop band edges and make some comments on a spatial evolution of the field passed through the periodic structure.

We start with considering characteristics of Bloch waves propagation through a periodic multilayer, whose unit cell has an inversion center. The unit cell is formed by three isotropic slabs, as illustrated by Fig. 1(c) of the main text. Spatial evolution of eigenmode fields over a period of the structure is described for each polarization by $2\times 2$ ABCD matrix $\overline{\overline P}$ acting on the field column $(H,E)^\textup{T}$ [where $H$ and $E$ are the tangential components of the magnetic and electric fields, respectively]. In case of the 1D photonic crystal the components of the ABCD matrix are as follows~\cite{Popov2018_OEMA_SW}
\begin{widetext}
\begin{eqnarray}\label{eq:TE_3_ABCDcoeff}
A&=&\cos(k_{z1}fd)\cos(k_{z2}[1-f]d) -\frac12\left(\frac{Z_{1}}{Z_2}+\frac{Z_2}{Z_1}\right)\sin(k_{z1}fd)\sin(k_{z2}[1-f]d),\nonumber\\
B&=&\frac{i}{Z_1}\left(\sin \left(k_{z1}fd \right) \cos \left(k_{z2}[1-f]d\right) +\frac12\left[\left(\frac{Z_{1}}{Z_2}-\frac{Z_2}{Z_1}\right)+\left(\frac{Z_{1}}{Z_2}+\frac{Z_2}{Z_1}\right)\cos \left( k_{z1}fd\right)\right]\sin \left(k_{z2}[1-f]d\right)\right),\nonumber\\
C&=&i Z_1\left(\sin \left(k_{z1}fd \right) \cos \left(k_{z2}[1-f]d\right) -\frac12\left[\left(\frac{Z_{1}}{Z_2}-\frac{Z_2}{Z_1}\right)-\left(\frac{Z_{1}}{Z_2}+\frac{Z_2}{Z_1}\right)\cos \left( k_{z1}fd\right)\right]\sin \left(k_{z2}[1-f]d\right)\right).
\end{eqnarray}
\end{widetext}
$Z_1$ and $Z_2$ are the wave impedances of either TE- or TM-polarized plane-waves in dielectric slabs of permittivities $\E_1$ and $\E_2$, respectively, $f$ is the fill fraction of $\E_1$, $d$ is the thickness of the unit cell, $k_{z1,2}=\sqrt{\E_{1,2}k_0^2-k_t^2}$, $k_t=k_0\sin\theta$, and $\theta$ is the angle of incidence.

\begin{figure}[tb]
	\includegraphics[width=0.99\linewidth]{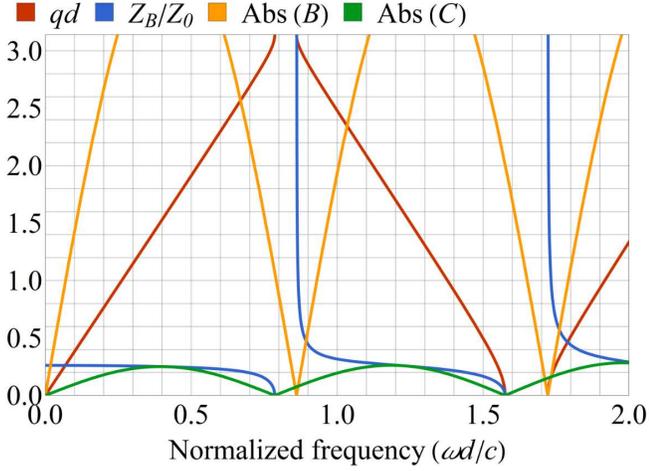}
	\caption{\label{fig:1S} Frequency behaviour of Bloch wavenumber $q$, wave impedance $Z_B$ and functions $B$ and $C$. Parameters: $\theta=0$,  $\E_1=2$, $\E_2=16$ and $f=0.1$.}
\end{figure}

A Bloch wave of the given polarization can be characterized by a wavenumber $q$ and wave impedance $Z_B$. They can be found by solving an eigenvalue problem for the transfer matrix $\overline{\overline P}$. The Bloch wavenumber $q$ satisfies the dispersion equation $\cos[qd]=A$ or, equivalently, $\sin[qd]=\sqrt{-BC}$, since $\overline{\overline P}$ is unimodular, that is, $A^2-BC=1$. 
The following explicit form of the dispersion equation can be found in the literature, e.g., in Ref.~\cite{yeh1977electromagnetic}:
\begin{eqnarray}
\label{eq:q}
q &=& \pm\frac1d\cos^{-1} \left[ \cos(k_{z1}d_1)\cos(k_{z2}d_2)-\frac12\left(\frac{Z_{1}}{Z_2}+\frac{Z_2}{Z_1}\right) \right. \nonumber\\
&\times& \left. \sin(k_{z1}d_1)\sin(k_{z2}d_2) \right].
\end{eqnarray}
When eigenvalues of $\overline{\overline P}$ are known, the wave impedances of Bloch waves can be found by solving the eigenvector equation
\begin{equation}
\left(\begin{array}{cc}
A&B\\
C&A
\end{array}
\right)\left(\begin{array}{c}
1\\
Z_B
\end{array}
\right)=e^{iqd}\left(\begin{array}{c}
1\\
Z_B
\end{array}
\right).
\end{equation}
After some algebra one arrives at a simple expression for the wave impedances:
\begin{equation}\label{eq:Z_B}
    Z_B=\pm\sqrt{\frac{C}{B}}.
\end{equation}
Signs plus and minus in Eqs.~\eqref{eq:q} and~\eqref{eq:Z_B} correspond to forward and backward Bloch waves, respectively. 
Edges of stop bands correspond to zeros of either function $B$ or $C$,  where wave impedance diverges or vanishes (see Eq.~\eqref{eq:Z_B} and Fig.~\ref{fig:1S}). Thus,  forward and backward Bloch waves are indistinguishable at stop band edges which can be treated as surfaces (owing to the axial symmetry) of degeneracy.

\begin{figure}[tb]
	\includegraphics[width=0.8\linewidth]{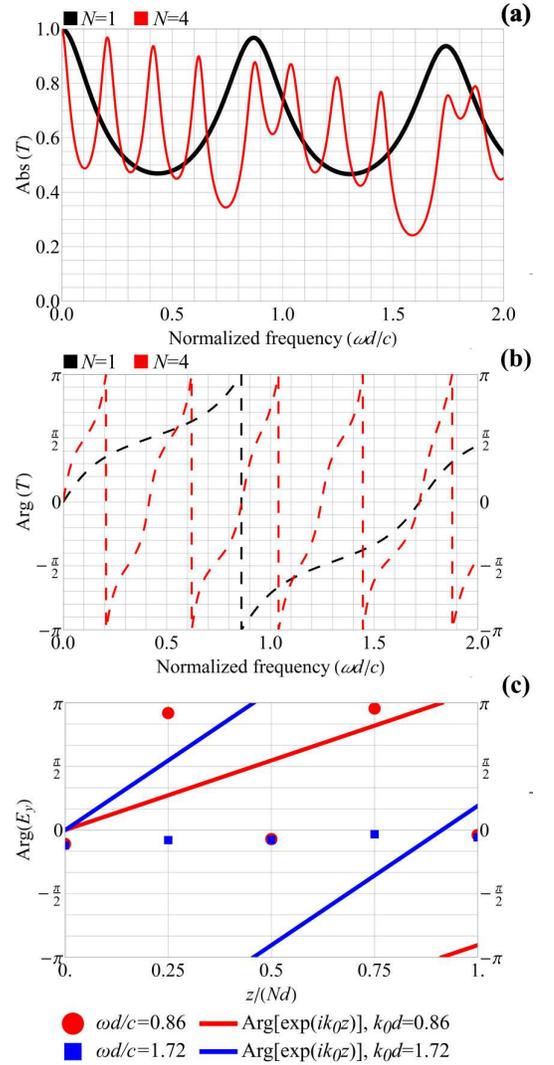}
	\caption{\label{fig:3S} Frequency dependence of (a) the absolute value and (b) phase of the  transmission coefficient through a 1D photonic crystal having $N$ unit cells. (c) Dependence  of the field phase on discrete  period's number $n$ of  the 1D photonic crystal having $N=4$ unit cells compared to the phase accumulation of a plane wave propagating through a vacuum region of equivalent thickness.  Parameters: normal incidence, $\E_a=1$, $\E_1=2(1+i 0.01)$, $\E_2=16(1+ i0.01)$ and $f=0.1$. Calculations were conducted in the assumption of $5\%$ random errors in the thicknesses of each dielectric layer.}
\end{figure}

In the general case, at a band gap edge, the transmission coefficient through $N$ unit cells of a 1D photonic crystal is not equal to $1$. It follows from Eq.~(1) in the main text which can be rewritten as  
\begin{equation}
T=\left[\cos(qNd)-i\frac{\frac{C}{B}+Z_a^2}{2Z_a}\sqrt{\frac{B}{C}}\sin(qNd)\right]^{-1}.
\end{equation}
Keeping in mind that $\sin(qd)=\sqrt{-BC}$, we find the transmission coefficient in the two limiting cases: 
\begin{equation}\label{eq:T_limit}
T\rightarrow\left\{
\begin{array}{cc}
    (1-NC/(2Z_a))^{-1}, & Z_B\rightarrow \infty  \\
    (1-NBZ_a/2)^{-1}, & Z_B\rightarrow 0.
\end{array}
\right.
\end{equation}
Albeit none of these limits equal unity, the transmission coefficient can approach $1$ arbitrarily close at a band gap edge, if the Brewster law is satisfied next to it (as in case of the examples demonstrated in the main text and Section S5 below).

Let us now discuss the spatial evolution of the field over periods of the structure. For the sake of brevity we omit the phrase ``over periods of the structure'' in what follows but implicitly assume it.  There are two principal situations corresponding to the incident plane-wave exactly matched with (i) the Brewster condition or (ii) a stop band edge. In the first case, the incident wave is impedance matched with the forward Bloch wave, which spatial evolution through the structure is then given by the phase factor $\exp[iqNd]$. Phases of both  magnetic and electric fields vary as $qNd$. In the other situation one has to turn to the ABCD matrix which at a stop band edge has $A=1$ and either $B=0$ ($C\neq0$) or $C=0$ ($B\neq 0$). 
The spatial evolution is given by the ABCD matrix acting on the field column $(1+R,Z_a(1-R))^\textup{T}$ at the input of the structure ($R$ is the reflection coefficient).
When $B=0$ the magnetic field does not change throughout the structure and has the phase $\textup{Arg}(H)=\textup{Arg}(1+R)$. Meanwhile, the phase profile of the electric field is given by the equation $\textup{Arg}(E)=\textup{Arg}(NC[1+R]+Z_a[1-R])$.
If $C$ is zero and $B$ is not, the electric field is  constant with the phase $\textup{Arg}(E)=\textup{Arg}(1-R)$ and the phase of the magnetic field changes as 
$\textup{Arg}(H)=\textup{Arg}(NBZ_a[1-R]+1+R)$.
Both situations become indistinguishable, when the Brewster condition occurs at a stop band edge.
It should be noted that exactly the same is true for a homogeneous ENZ slab which is treated by means of the similar analytical technique and has the similar physics.

To conclude we would like to note that the main results of this section are obviously applicable for any periodic structure with a symmetric unit cell, not only the multilayer.


\section*{S3: Analysis of  tolerances}

In this section we discuss tolerances with respect to small changes in the frequency and structure's geometry related to the 1D photonic crystal considered in the main text.    
Requirements to the design of the unit cells are quite flexible as one does not have to satisfy exact mathematical conditions. However, there is a general rule that a photonic crystal should be formed of alternating high- and low-permittivity dielectric layers, with the fill fraction of the low-permittivity material being small. 
Influence of the dielectric losses and moderate ($5\%$) random errors of the layers thicknesses is demonstrated in Fig.~\ref{fig:3S}. The losses do not destroy the effect of zero phase delay, but obviously reduce the transmission amplitude at the resonance frequencies as clearly seen from comparison of Figs.~3(c)--(e) with Figs.~\ref{fig:3S}(a), (b). The widths and positions of the peaks only slightly change compared with the lossless multilayer. The random errors in addition to the losses affect mainly the phase accumulation, but the latter is still quite small, as shown in Fig.~\ref{fig:3S}(c).

\section*{S4: Analysis of group delay}

\begin{figure}[tb]
	\includegraphics[width=0.8\linewidth]{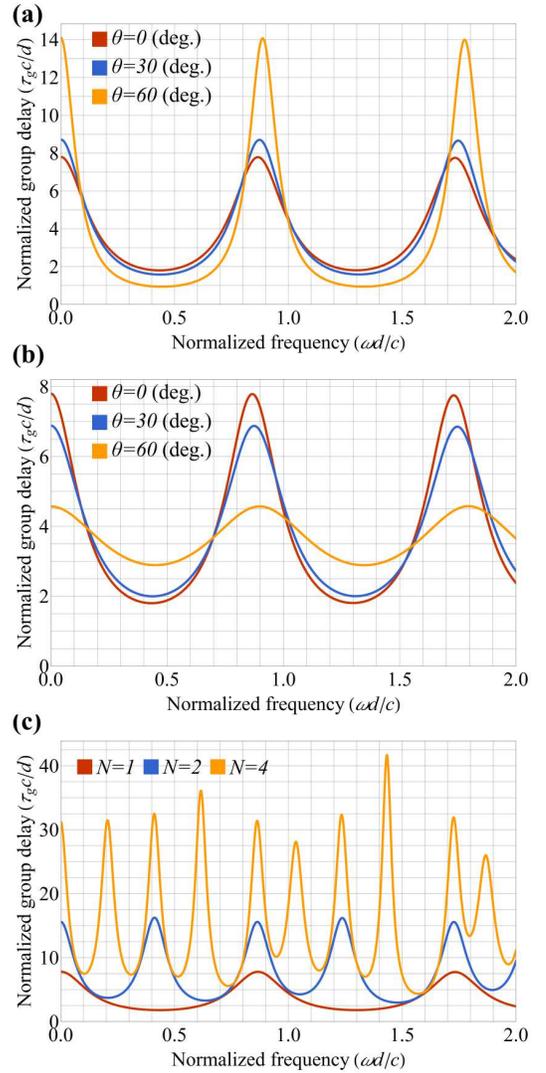}
	\caption{\label{fig:2S} Frequency dependence of normalized group delay. (a), (b)  Through a single unit cell for different incidence angles in case of (a) TE and (b) TM polarizations. (c) Through 1D photonic crystal having different number of unit cells $N$ versus normalized frequency in case of  normal incidence. Other parameters: $\E_a=1$, $\E_1=2$, $\E_2=16$ and $f=0.1$. }
\end{figure}
\begin{figure*}[tb]
\centering
	\includegraphics[width=0.99\linewidth]{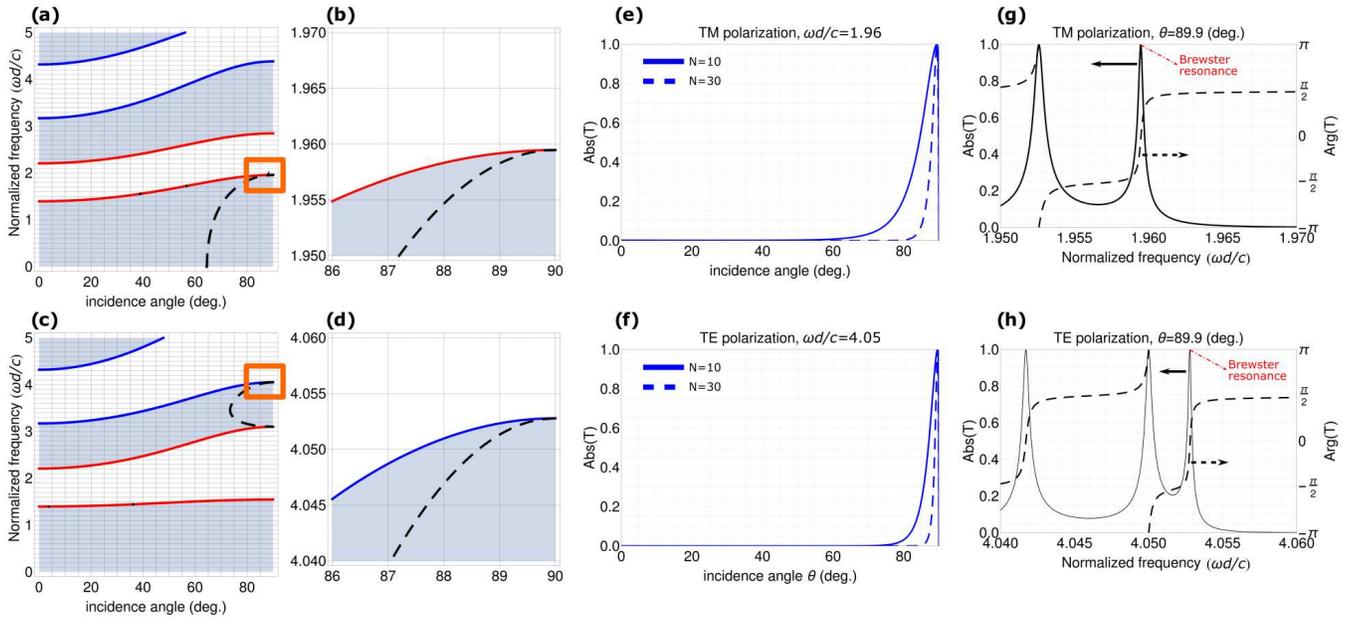}
	\caption{\label{fig:4S} (a), (c) Band gap structures of a photonic crystal illustrated in Fig. 1 (b): (a) TM polarization and (c) TE polarization.  (b), (d) Zooming of the boxed areas in Figures (a) and (c), respectively. The Brewster condition is satisfied at the black dashed curves. Shaded regions represent propagation bands,  red and blue curves represent the edges of stop bands where $q=\pi/d$ and $q=0$, respectively. (e), (f) Angular dependence of the transmission coefficient through  slabs of photonic crystals having $10$ and $30$ unit cells. (g), (h) Frequency dependence of the transmission coefficient  through the $30$ unit cells photonic crystal. In all figures  the material  parameters are $\E_1=2$, $\E_2=16$, and the volume fraction of the low-index material is $f=0.9$. }
\end{figure*}
\begin{figure}[tb]
\centering
	\includegraphics[width=0.99\linewidth]{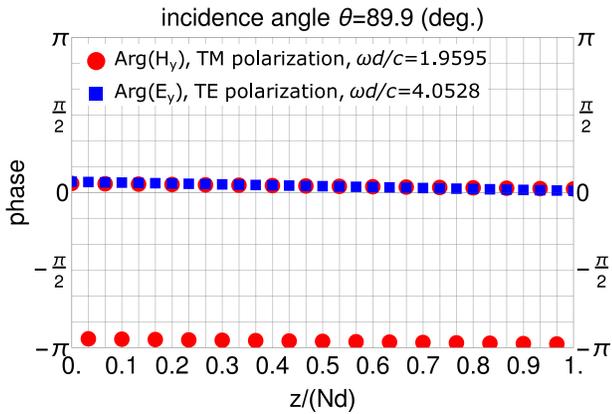}
	\caption{\label{fig:5S} Distribution of the field phase throughout the photonic crystal having $30$ unit cells, cases of TM ($k_0d=1.9595$) and TE ($k_0d=4.0528$) polarizations. The points represent the field phase at the interfaces of unit cells.  The structure parameters are $\E_1=2$, $\E_2=16$, and $f=0.9$. }
\end{figure}

In the main text we deal only with monochromatic plane waves. 
Generally, a spectral composition of such waves (signal) may propagate with  significant distortions due to a complex frequency response of the propagation medium. 
From Figs.~3~(c)~--~(e) one can see that  the transfer function of the considered 1D photonic crystal (namely, the transmission coefficient $T$) significantly depends on the frequency.
In order to estimate possible signal distortion due to the nonlinear (with respect to the frequency) phase response  we consider group delay   $\tau_g$ defined as $\partial\textup{Arg}[T]/\partial\omega$. Group delay at the frequency $\omega$ can be understood as the time it takes for a signal with a narrow spectrum (centered around $\omega$)  to traverse the structure~\cite{Oppenheim1996}. Figure~\ref{fig:2S} shows the normalized  group delay through the 1D photonic crystal considered in the main text as a function of frequency. Results for different polarizations and incidence angles are presented by Figs.~\ref{fig:2S} (a) and (b) for a single unit cell. The maxima of the group delay (except the one at $\omega=0$)  correspond to the top edges of band gaps (in coherence with the well-known slow light phenomenon~\cite{Dowling1994}), where we expect to observe full transmission without phase delay.  Notably, in case of TE polarization the magnitude of maxima increases with the incidence angle, while it decreases for TM polarization. 
Group delay increases and additional maxima appear, when one adds unit cells to the photonic crystal slab, see Fig.~\ref{fig:2S} (c). 
Overall, we can conclude that only signals of small bandwidth can be used for transferring information through such a photonic crystal possessing ENZ properties.

\section*{S5: Another example with 1D photonic crystal: \\ENZ-like spatial frequency filtering}
\label{app:e}

A thick epsilon-near-zero slab acts as an effective filter of spatial frequencies transmitting only normally incident waves.
Although in case of 1D photonic crystals it is difficult to get such a functionality for the normally incident wave, sharp transmission resonances are well accessible for grazing incident waves, when the wavevector of the incident wave is almost parallel to the photonic crystal interfaces. To that end, we study a  multilayer composed by alternating low- and high-permittivity dielectric layers keeping the filling fraction $f = d_1/d$ of the low-permittivity dielectric \textit{high}. 
Figures~\ref{fig:4S}(a) and (c) depict the band gap structure for TM and TE polarizations. The Brewster condition is met at the black dashed curves. 
Blue and red curves correspond to the edges of stop bands $q=0$ and $q=\pi/d$, respectively. By magnifying the red boxed areas we clearly see from Figs.~\ref{fig:4S}(b) and (d) that the Brewster effect occurs at a stop band edge for both polarizations (but at different frequencies), when  the incidence angle $\theta$ approaches $90$~degrees. In Figs.~\ref{fig:4S}(e) and (f) we plot the angular dependence of the transmission coefficient at the corresponding frequencies for photonic crystals composed of $10$ (solid curve) and $30$ (dashed curve) unit cells. Since the whole range of the incidence angles apart from a vicinity of $90$~degrees corresponds to a stop band, highly selective transmission is observed. 
The transmission resonances do not depend on the total thickness $Nd$ of the multilayer that is they are Brewster's resonances. 
When the Brewster condition is satisfied at a stop band edge, the phase of the transmission coefficient vanishes. It can be seen from the frequency dependence of the transmission coefficient around the stop band edge frequency (the incidence angle is $89.9$ degrees), shown in Figs.~\ref{fig:4S}(g) and (h), where one can actually recognize both the Brewster and the Fabry-Perot resonances.  The Brewster resonances appear right at the stop band edge.

Full power transmission occurs also at Fabry-Perot resonances, where 
$q Nd=\pi m$ ($m$ is an integer number), in this case the transmission coefficient given by Eq.~(1) in the main text is $T=1/\cos(\pi m)$. The Fabry-Perot resonances are seen in Figs.~\ref{fig:4S}(g) and (h) at frequencies below the Brewster peak at the stop band edge.  Interestingly, for even  $m$ we have  $T=1$ but the phase accumulation is $2\pi m$.  
As it is noted above, the Brewster effect is independent on $Nd$, and there is no phase accumulation in the full transmission regime, which makes it fundamentally different from  the Fabry-Perot  resonances. 

Absence of the phase accumulation over the full length of the photonic crystal with $30$ unit cells is demonstrated by Fig.~\ref{fig:5S}. Field phases at the interfaces of unit cells are shown at the incidence angle of $89.9$ degrees. In case of the TM polarization the stop band edge of interest corresponds to $q=\pi/d$ ($k_0d \approx 1.96$), and  there is no phase accumulation between discrete points $z_m = 2 m d$ or $z_m = (2m+1) d$ ($m$ is an integer number).
For the TE polarization the Brewster condition is satisfied at the stop band edge $q=0$, when the frequency is approximately $k_0d=4.05$, and zero phase accumulation is observed after every period.

In order to conclude this section let us note that the results represented by Figs.~\ref{fig:4S} and \ref{fig:5S} look nicer than those demonstrated in the main text. It is due to the fact that the wave impedances of incident plane-waves diverge for TE polarization and vanish in case of TM one at ninety-degrees incidence. Meanwhile, the wave impedances of Bloch waves behave the same at stop band edges (see Section S2). Thus, the Brewster condition is satisfied \textit{exactly} at the considered stop band edges.

	\bibliography{bib}
\end{document}